\documentclass{article}
\usepackage{arxiv}
\renewcommand{\shorttitle}{Physics-guided machine learning for saturation control in heap leaching}

\usepackage[utf8]{inputenc} 
\usepackage[T1]{fontenc}

\usepackage{amsmath}
\usepackage{amssymb}        
\DeclareMathOperator*{\argmin}{arg\,min}
\usepackage{graphicx}       
\graphicspath{{Pics/}}      
\usepackage{float}
\usepackage{booktabs}       
\usepackage{array}
\usepackage{enumitem}
\usepackage{xcolor}
\usepackage[numbers,sort&compress]{natbib}
\usepackage{url}            
\usepackage{hyperref}       
\hypersetup{colorlinks=true, linkcolor=blue!50!black, citecolor=blue!50!black, urlcolor=blue!50!black}
\usepackage[capitalize]{cleveref}
\usepackage{setspace} 

\newcommand{\sub}[1]{#1}

\title{Physics-guided machine learning for saturation control across variable hydraulic conditions in heap leaching}

\author{
  Sultan Alhamdan \\
  Department of Chemical Engineering \\
  University College London \\
  London, United Kingdom \\
  \And
   \href{https://orcid.org/0000-0003-3541-3692}{\includegraphics[scale=0.06]{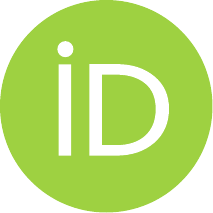}}\hspace{1mm} Daniel Navia \\
  Department of Chemical and Environmental Engineering \\
  Universidad T\'ecnica Federico Santa Mar\'{\i}a \\
  Valpara\'{\i}so, Chile \\
  \And
  \href{https://orcid.org/0000-0002-7717-0556}{\includegraphics[scale=0.06]{orcid.pdf}}\hspace{1mm} Paulina Quintanilla\thanks{Corresponding author. Email: \texttt{p.quintanilla@ucl.ac.uk}} \\
  Department of Chemical Engineering \\
  University College London \\
  London, United Kingdom \\
}
\hypersetup{
pdftitle={Physics-guided machine learning for saturation control across variable hydraulic conditions in heap leaching},
pdfsubject={Physics-guided machine learning for saturation control across variable hydraulic conditions in heap leaching},
pdfauthor={Sultan Alhamdan, Daniel Navia, Paulina Quintanilla},
}
\begin{document}
\raggedbottom
\maketitle

\begin{abstract}
Heap-leaching hydraulics vary between regions of an ore bed, motivating fast prediction models that accommodate different material properties. This study develops a parameter-conditioned physics-guided machine-learning (PGML) surrogate and demonstrates its use in simulated saturation control. A dataset of 4096 numerical Richards-equation trajectories was divided into 2867 training, 614 validation and 615 evaluation cases. The network predicts hourly saturation profiles from the current profile, irrigation rate and four hydraulic parameters, and is trained with an approximate Gardner mass-balance penalty. A daily predictive controller compared 200 constant irrigation rates over a 24-hour horizon. Across the evaluation cases, the median full-cycle tracking deviation was 4.9\% of a 0.70 saturation target, equal to that of a tuned proportional-integral (PI) controller, while the PGML controller reached the target without overshoot in the median and applied approximately 4\% less irrigation. Moderate physics guidance reduced long-horizon prediction error by approximately 30\% relative to data-only training and reduced irrigation-bound activity, but slightly worsened closed-loop tracking. For a matched prediction workload, the surrogate was approximately 70 times faster than the numerical solver, and batched evaluation cost under 2~ms per column for 5000 columns. The results support a reusable physics-guided surrogate for large-scale saturation control in simulation.

\end{abstract}

\keywords{Heap leaching \and Saturation control \and Physics-guided machine learning \and Model predictive control \and Irrigation \and Hydraulic variability}

\section{Introduction}
\label{sec:intro}

Copper demand is expected to keep rising with electrification and the energy transition \citep{InternationalEnergyAgency2025}, while the average grade of mined copper ores continues to decline \citep{Calvo2016}. Processing routes that recover metal economically from low-grade ores are therefore increasingly important, and heap leaching is one of the most established. In this process, crushed and often agglomerated ore is stacked on an impermeable pad and irrigated with a leaching solution. The solution percolates through the bed, dissolves the metal and is collected at the base as pregnant leach solution (PLS) \citep{Petersen2016,McBride2018}. Compared with milling and flotation, heap leaching has low capital and operating costs, which makes it attractive for low-grade copper ores \citep{Petersen2016}.

The performance of a heap depends strongly on how the leaching solution is distributed within the ore bed. Poor wetting and preferential flow can leave parts of the bed with little contact with the solution, even when the total irrigation rate is at its design value \citep{Liu2017,McBride2018}. At the same time, metal recovery responds slowly to changes in irrigation and is usually measured at the PLS outlet only daily or weekly, which delays any corrective action \citep{Olivares2025}. The moisture content of the bed, expressed as the degree of saturation, responds much faster and is closely linked to the liquid--solid contact that drives dissolution. Regulating saturation through the irrigation rate therefore offers a faster way of acting on the hydraulic state of the heap \citep{Olivares2025}. Because a heap is irrigated by many drippers, a practical approach divides it into independently irrigated control volumes. Each control volume is represented as a one-dimensional column, in which Richards' equation describes the vertical movement of moisture \citep{Richards1931,Cariaga2015,Winarko2023,Gonzlez2026}.

Several strategies have been proposed for setting the irrigation rate. Optimisation methods have been used to generate irrigation plans \citep{Jlvez2025} and to determine optimal infiltration conditions \citep{Eppink2026}. Feedback strategies, in which the irrigation rate is updated from measurements of the heap, have also been explored. \citet{Canales2024} trained a deep reinforcement-learning agent to schedule irrigation, and \citet{Olivares2025} developed a nonlinear model predictive control (NMPC) framework to regulate saturation in a single column. In NMPC, a process model predicts how the column will respond to different irrigation trajectories, an optimiser selects the best trajectory, and only the first decision is applied before the procedure is repeated with new measurements. \citet{Olivares2025} combined this procedure with online identification of the hydraulic parameters and showed that moisture measurements are needed when the model differs from the process. Their calculations took a few seconds per decision for one column, well within a daily control interval.

Extending model-based control to many columns requires both hydraulic characterisation and repeated prediction. A parameter-conditioned surrogate can share one set of network weights across columns, reducing the need for separate surrogate training. It does not remove the need to determine each column's hydraulic parameters: a mechanistic model likewise uses a common equation structure with column-specific parameters. The potential computational benefit is therefore in repeated forward prediction once those parameters are supplied. Its control performance should also be compared with simpler feedback rules, which are easier to implement and maintain. Physics-guided machine learning (PGML) offers a route to such a model. A PGML model is a data-driven model, typically a neural network, whose training also takes into account known physical relations, such as conservation laws \citep{Karniadakis2021,Willard2023}. Once trained, it can be evaluated much faster than a numerical solution of the governing equations. For unsaturated flow, physics-informed networks have been used to infer constitutive relations and hydraulic properties from moisture data \citep{Bandai2021,Depina2022}. In heap leaching, \citet{Bravo-Gutirrez2026} combined a phenomenological flow model with a neural network that represents the dissolution chemistry, and \citet{Olivares2025} identified physics-informed neural networks as a possible way of extending saturation control. However, machine-learning models in mineral processing must be evaluated carefully, because their performance outside the training conditions is not guaranteed \citep{Estay2023}.

This paper develops a parameter-conditioned PGML surrogate for hourly saturation prediction and examines prediction--control trade-offs in simulated heap-leaching columns. The network learns numerical Richards-equation trajectories, with an approximate water mass-balance penalty based on Gardner's relations \citep{Gardner1958}. A daily controller evaluates constant-rate predictions over 24 hours. The contribution is a shared surrogate across sampled hydraulic conditions and a comparison of its prediction, tracking and computation characteristics with data-only training and alternative controllers. All process outcomes are simulations. \Cref{sec:process} defines the problem, \Cref{sec:methods} presents the methodology, \Cref{sec:results} discusses the results and limitations, and \Cref{sec:conclusions} gives the conclusions.

\section{Process description and control problem}
\label{sec:process}

\begin{figure}[H]
    \centering
    \includegraphics[width=0.70\linewidth]{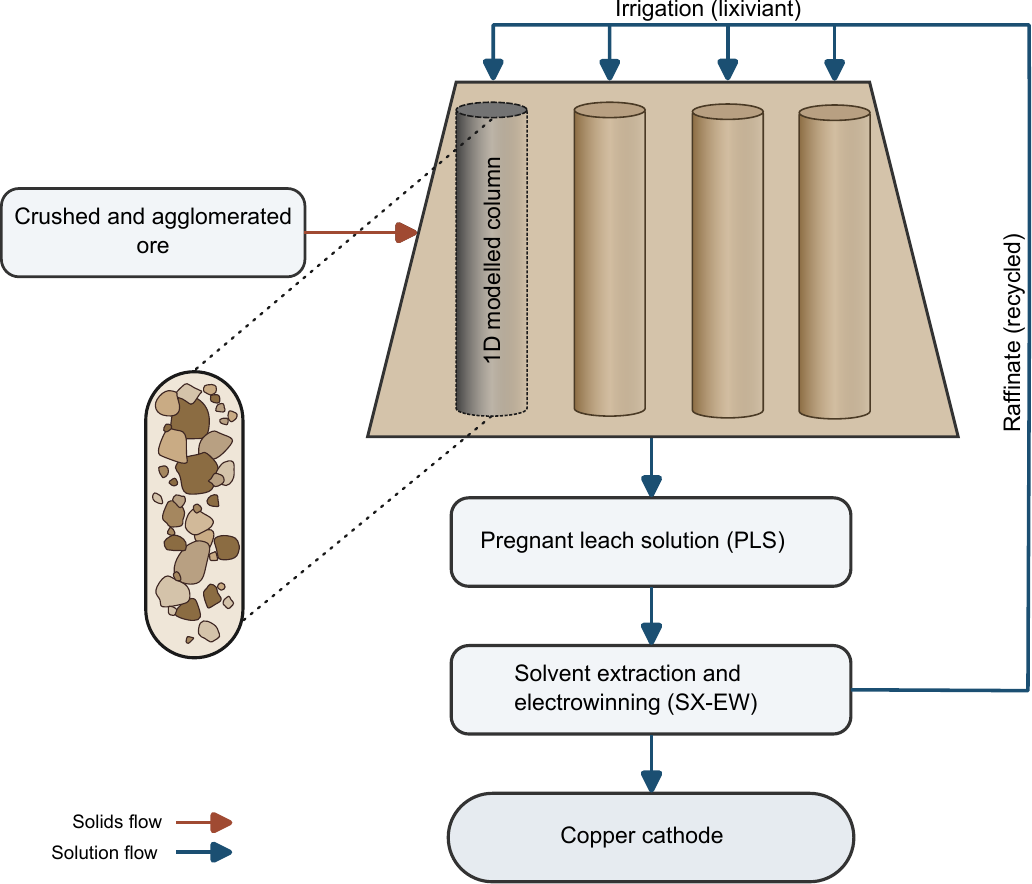}
    \caption{Copper heap-leaching circuit and the column representation used for local saturation control. Each independently irrigated column receives a prescribed surface flux; its effective hydraulic properties describe the vertical flow model.}
    \label{fig:heap}
\end{figure}

In a copper heap, crushed and agglomerated ore is stacked to a height of several metres on a lined pad. Drippers or sprinklers distribute the leaching solution over the surface, and the solution flows downwards through the unsaturated bed under the combined action of gravity and capillary forces. The PLS collected at the base is sent to the recovery plant, and the barren solution returns to the irrigation system (\Cref{fig:heap}). Part of the pore space is filled with solution and the rest with air. The degree of saturation, $S_e=\theta/\theta_s$, describes the filled fraction, where $\theta$ is the volumetric moisture content and $\theta_s$ its value when the pores are full; the residual moisture content is taken as zero.

Following previous work \citep{Winarko2023,Olivares2025,Gonzlez2026}, the heap is divided into control volumes, each irrigated by its own dripper and represented as a vertical column (\Cref{fig:heap}). Each column has its own hydraulic properties, which describe how quickly the solution enters the bed, spreads within it and drains from it. Columns exchange no solution laterally, so the irrigation of each column can be decided separately. This assumption isolates the effect of differences in hydraulic properties between columns, which is the focus of the present work.

The control problem for each column is to choose the irrigation rate so that the saturation profile reaches a target value and stays there. In control terms, the irrigation rate is the manipulated variable, the saturation profile is the controlled variable and the target value is the set-point. A set-point of $S_e^{SP}=0.70$, one of the values considered by \citet{Olivares2025}, was used throughout. The irrigation rate is updated once a day and held constant between updates. In feedback control, each decision uses the latest measurement of the column, so that deviations from previous days are corrected. In practice, the moisture profile along a column can be measured with electrical resistivity tomography (ERT) \citep{Olivares2025}.

\begin{figure}[!htbp]
    \centering
    \includegraphics[width=0.85\linewidth]{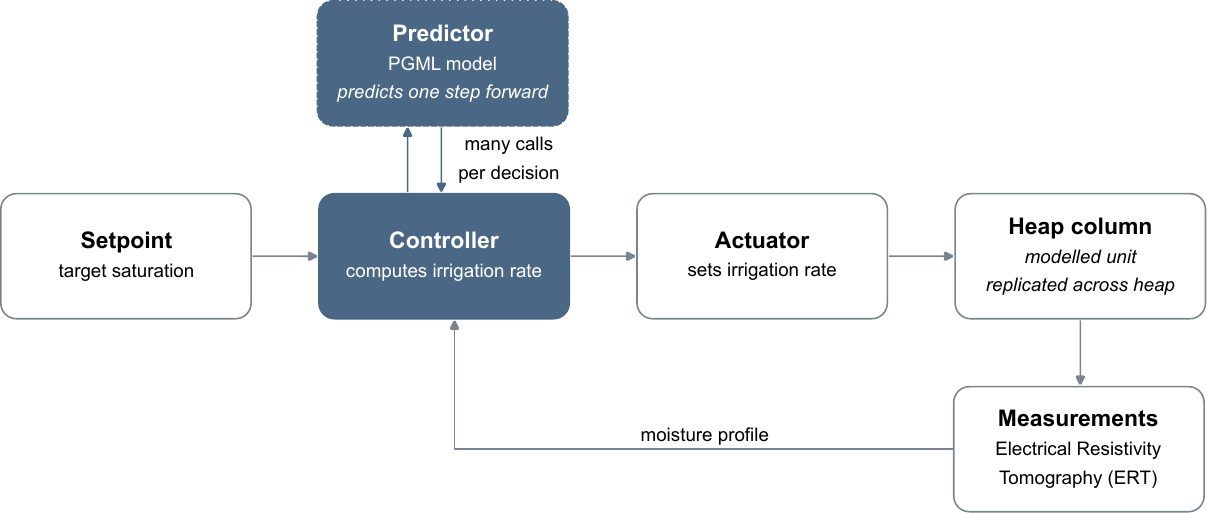}
    \caption{Closed-loop irrigation control of one column. Each day, the controller receives the measured saturation profile, uses the PGML model to predict the response to candidate irrigation rates, and sends the selected rate to the irrigation system. The same structure is replicated for every column of the heap.}
    \label{fig:loop}
\end{figure}

Two broad families of feedback controllers can solve this problem (\Cref{fig:loop}). Rule-based controllers, such as the widely used proportional--integral (PI) controller, calculate the irrigation rate directly from the difference between the target and the measured saturation. They are simple to implement, but they do not anticipate how the column will respond. Model predictive control (MPC) instead uses a model of the column to predict its response to candidate irrigation rates over a future period, called the prediction horizon, and selects the rate whose predicted outcome is best. The selected rate is applied for one day, the column is measured again and the procedure is repeated; this repetition is known as a receding horizon. Because MPC requires many model evaluations at each decision, a fast and reusable model, such as the PGML surrogate developed here, becomes valuable when many columns must be controlled.

\section{Methodology}
\label{sec:methods}

The proposed framework has three parts: a numerical model of a leaching column, which generates the training data and serves as the reference process in all tests (\Cref{sec:model}); a PGML surrogate trained on simulations across a range of hydraulic conditions (\Cref{sec:sampling,sec:pgml}); and a predictive controller that uses the surrogate to select the daily irrigation rate (\Cref{sec:control}). \Cref{sec:baselines_method} describes the controllers used for comparison, and \Cref{sec:protocol} defines the case studies and performance measures.

\subsection{Hydraulic model of a leaching column}
\label{sec:model}

Moisture movement in each column is described by Richards' equation, which combines mass conservation with Darcy's law for unsaturated flow \citep{Richards1931}. In moisture-content form, with depth $z$ increasing downwards,

\begin{equation}\label{eq:richards1d}
    \frac{\partial \theta}{\partial t} = \frac{\partial}{\partial z}\left[ D(\theta)\,\frac{\partial \theta}{\partial z} \right] - \frac{\partial K(\theta)}{\partial z}
\end{equation}
where $K$ is the hydraulic conductivity and $D$ the moisture diffusivity. Both depend on the moisture content through constitutive relations. The simulator uses the van Genuchten Modified (VGM) relations of \citet{Cariaga2015}. With zero residual moisture, the effective saturation and the conductivity are

\begin{equation}\label{eq:vg}
    S_e = \frac{\theta}{\theta_s} = \left[ 1 + (\alpha|h|)^{n} \right]^{-m}, \qquad m = 1 - \frac{2}{n}, \quad n > 2,
\end{equation}
\begin{equation}\label{eq:vgm_k}
    K(S_e) = K_{s}\,S_e^{\delta}, \qquad \delta = 3 + \frac{2}{n-2},
\end{equation}
where $h$ is the pressure head, $K_s$ the saturated hydraulic conductivity, and $\alpha$ and $n$ set the shape of the retention curve. Through the exponent $\delta$, the parameter $n$ also controls how quickly the conductivity falls as the bed dries. The diffusivity follows as $D(\theta)=K(\theta)/(\mathrm d\theta/\mathrm dh)$. Because $\theta_s$ is constant within each column, the equation can be written in terms of saturation:

\begin{equation}\label{eq:se_system}
    \frac{\partial S_e}{\partial t} = \frac{\partial}{\partial z}\left[ D(S_e)\,\frac{\partial S_e}{\partial z} \right] - \frac{1}{\theta_s}\,\frac{\partial K(S_e)}{\partial z}.
\end{equation}
The initial saturation is uniform, $S_e(z,0)=S_{e,\mathrm{init}}$. At the surface ($z=0$), the irrigation rate $R$ sets the incoming flux, $-\theta_sD(S_e)\,\partial S_e/\partial z+K(S_e)=R$. At the base ($z=H$), a free-drainage condition, $\partial S_e/\partial z=0$, gives an outflow $q(H)=K(S_e(H))$.

The equations were solved with a modified Crank--Nicolson scheme on 221 nodes over a height of $H=550$~cm ($\Delta z=2.5$~cm), with a time step of one hour over a 25-day cycle. Each hourly step applied 20 fixed-point corrections, and each linear system was solved with the Thomas algorithm \citep{Cariaga2015,Davis1984}. The geometry and the reference parameter values follow \citet{Cariaga2015}, who estimated them for a copper heap. To keep the constitutive expressions finite, moisture values were clipped just inside their physical bounds. Two numerical diagnostics were recorded: bound violations before clipping, which flag numerical difficulties, and the residual of the final correction. This simulator produced the training data and served as the reference process in all control tests, so all results are measured against this implemented numerical model.

A simpler constitutive model was used for the physics term of the surrogate. For $h\leq0$, Gardner's exponential relations \citep{Gardner1958,Tracy2010,Olivares2025} give

\begin{equation}\label{eq:gardner}
    S_e(h)=e^{\alpha_g h/\beta},\qquad K_G(h)=K_s e^{\alpha_g h},
\end{equation}
so that
\begin{equation}\label{eq:gardner_se}
    K_G(S_e)=K_s S_e^\beta,\qquad
    D_G(S_e)=\frac{\beta K_s}{\alpha_g\theta_s}S_e^{\beta-1}.
\end{equation}
Setting $\beta=\delta$ makes the Gardner conductivity equal to the VGM conductivity in \Cref{eq:vgm_k}, whereas the Gardner retention curve and diffusivity only approximate the VGM ones. The Gardner form was chosen because it gives a simple, differentiable expression for the water mass balance, which is needed during training. The training data therefore describe the VGM behaviour, and the Gardner mass balance adds an approximate physical constraint. Appendix~\ref{app:residual} gives the discrete fluxes and boundary treatment.

\subsection{Simulated dataset}
\label{sec:sampling}

The study samples a prescribed hydraulic range around a literature reference case, rather than an empirically established distribution for an entire heap. The dataset was generated by varying four hydraulic properties, $K_s$, $n$, $\alpha$ and $\theta_s$, and the initial saturation $S_{e,\mathrm{init}}$ within the bounds in \Cref{tab:envelope}. Scrambled Sobol sampling \citep{Sobol1967} spread 4096 parameter sets evenly across this five-dimensional range. The ranges were first checked with the numerical model to ensure that every combination could be simulated at the maximum irrigation rate, $R/K_s=0.30$. At the nominal $K_s=170$~cm/day, typical industrial irrigation rates of 5--20~L/m$^2$/h correspond to $R/K_s\simeq0.07$--0.28 \citep{Petersen2016}.

\begin{table}[H]
  \centering
  \caption{Parameter bounds used to sample the simulated dataset. The nominal value is the reference parameter set of \citet{Cariaga2015}. The operational irrigation cap is $R/K_s\leq0.30$.}
  \label{tab:envelope}
  \footnotesize
  \setlength{\tabcolsep}{6pt}
  \renewcommand{\arraystretch}{1.15}
  \begin{tabular}{@{} l l l l @{}}
    \toprule
    \textbf{Symbol} & \textbf{Quantity} & \textbf{Nominal value} & \textbf{Range} \\
    \midrule
    $K_s$ [cm/day]      & saturated hydraulic conductivity  & 170    & 130 -- 232     \\
    $n$ [--]            & van Genuchten shape parameter     & 2.267  & 2.20 -- 3.00   \\
    $\alpha$ [1/cm]     & van Genuchten retention parameter & 0.035  & 0.025 -- 0.042 \\
    $\theta_s$ [--]     & saturated moisture content        & 0.330  & 0.270 -- 0.388 \\
    $S_{e,\mathrm{init}}$ [--] & initial column saturation    & 0.424  & 0.40 -- 0.63   \\
    \bottomrule
  \end{tabular}
\end{table}

Each parameter set was simulated for 25 days under its own irrigation schedule. Because the surrogate must learn how saturation responds to changes in irrigation, the schedules contained daily changes in the irrigation rate \citep{Herkersdorf2024}. For each parameter set, two additional Sobol coordinates defined a target irrigation fraction $q_f\in[0.10,1]$ and a log-amplitude $\sigma\in[0,1]$. With $\bar R=q_f(0.30K_s)$ and $u_d$ drawn uniformly from $[-1,1]$, the daily irrigation rate was

\begin{equation}\label{eq:irrigation_excitation}
 R_0=R_{\mathrm{min}},\qquad
 R_d=\operatorname{clip}\!\left(\bar R e^{\sigma u_d},\;
 R_{\mathrm{min}},\;\min(0.30K_s,2R_{d-1})\right),
\end{equation}
where $R_{\mathrm{min}}=R_{\mathrm{scale,min}}\simeq0.162$~cm/day is the lowest rate in the training data, which is also the lower bound used by the controller (\Cref{sec:control}). The rate could at most double from one day to the next and never fell below this minimum. Each run stored 600 end-of-hour saturation profiles, which gave 599 hourly transitions, each paired with the irrigation rate applied during that hour.

The 4096 runs were split into 2867 training, 614 validation and 615 evaluation cases (the latter are labelled \texttt{test} in the released code). Whole runs were assigned to each split, so neighbouring hourly states from one trajectory never appeared in different splits. The training split contained about 1.72 million transitions, and input scalings were fitted on it. Validation cases informed checkpoint selection, physics-weight comparison and PI tuning (\Cref{sec:baselines_method}). Evaluation cases were excluded from network-weight fitting, but informed batch-size selection (Appendix~\ref{app:batch}), and 74 of them were among the 512 runs used for Gardner calibration (\Cref{sec:closure}). Sampling and splitting used a fixed random seed.

\subsection{Physics-guided surrogate model}
\label{sec:pgml}

The surrogate replaces the numerical model inside the controller. It is a multilayer perceptron, a standard feed-forward neural network, that predicts the saturation profile one hour ahead:

\begin{equation}\label{eq:onestep}
    \hat{S}_e^{k+1} = f\big(S_e^k,R^k,K_s,\alpha,n,\theta_s\big),
\end{equation}
where $k$ is the hourly step, $R^k$ the irrigation rate during that hour and the circumflex denotes a prediction. The current profile describes the state of the column, and the four hydraulic parameters describe the material. Because these parameters are inputs, one set of network weights accepts different column parameters within the sampled range: a new column requires only its parameters and current profile, not a new network. Predictions over longer periods are obtained by applying \Cref{eq:onestep} recursively, feeding each prediction back as the next input.

\begin{figure}[!htbp]
    \centering
    \includegraphics[width=0.85\linewidth]{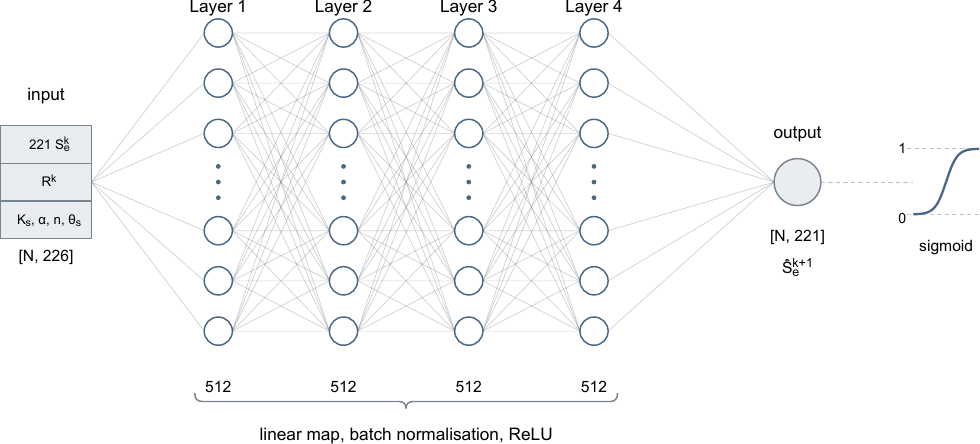}
    \caption{PGML network: 226 inputs pass through four hidden layers of width 512 to give 221 saturation predictions. Each hidden layer applies a linear map, batch normalisation and ReLU, and a sigmoid keeps the output within $[0,1]$. $R^k$ is the irrigation rate during the predicted hourly step $k$ (\Cref{eq:onestep}).}
    \label{fig:network}
\end{figure}

The profile is given at the 221 nodes of the numerical grid, so the network has 226 inputs (221 saturation values, the irrigation rate and four parameters) and 221 outputs. The irrigation and hydraulic inputs are scaled to $[0,1]$. Four hidden layers of 512 units each apply a linear map, batch normalisation and a rectified linear (ReLU) activation, and a sigmoid output keeps the predicted saturation within $[0,1]$ (\Cref{tab:arch,fig:network}). The network was trained with the Adam optimiser \citep{Kingma2015}, with the learning rate scaled to the batch size \citep{Hoffer2017}.

\begin{table}[H]
  \centering
  \caption{PGML network architecture and training configuration.}
  \label{tab:arch}
  \footnotesize
  \setlength{\tabcolsep}{7pt}
  \renewcommand{\arraystretch}{1.2}
  \begin{tabular}{@{} l l @{\hspace{3em}} l l @{}}
    \toprule
    \multicolumn{2}{c}{\textbf{Architecture}} & \multicolumn{2}{c}{\textbf{Training}} \\
    \midrule
    input / output dimension & 226 / 221    & optimiser                & Adam \\
    hidden layers             & 4            & learning rate            & 0.004 \\
    hidden width               & 512          & batch size               & 4096 \\
    hidden block               & BatchNorm, ReLU & physics weight $\lambda$ & 0.4 \\
    output activation          & sigmoid      & epochs (warm-up)         & 150 (5) \\
    trainable parameters       & 1.0 million  & LR decay                 & halve on plateau, patience 10 \\
    \bottomrule
  \end{tabular}
\end{table}

\subsubsection{Training loss}
\label{sec:loss}

Training minimises a weighted combination of two errors. The data loss $\mathcal L_d$ is the mean squared difference between predicted and simulated saturation. The physics loss $\mathcal L_p$ is the mean squared residual of a discrete water mass balance, written with the Gardner relations and evaluated at the prediction (Appendix~\ref{app:residual}). In other words, $\mathcal L_p$ is small when the predicted change in stored water in each layer matches the water that enters and leaves that layer during the hour. Because the two losses have very different magnitudes, each is divided by its own current value before weighting; this value is held constant during differentiation by the stop-gradient operator $\mathrm{sg}(\cdot)$ \citep{Bischof2025}:

\begin{equation}\label{eq:scaling}
\hat{\mathcal{L}}_d = \frac{\mathcal{L}_d}{\max(\mathrm{sg}(\mathcal{L}_d),\eta)}, \qquad
\hat{\mathcal{L}}_p = \frac{\mathcal{L}_p}{\max(\mathrm{sg}(\mathcal{L}_p),\eta)}, \qquad
\mathcal{L} = (1-\lambda)\,\hat{\mathcal{L}}_d + \lambda\,\hat{\mathcal{L}}_p,
\end{equation}
Here $\eta=10^{-30}$ avoids division by zero, and $\lambda\in[0,1]$ is the physics weight: $\lambda=0$ gives purely data-driven training and $\lambda=1$ purely physics-based training. After scaling, $\lambda$ sets the balance between the relative improvements in the two losses. Because the scaled loss always stays close to one, the unscaled validation data error was used to select the best network and to adjust the learning rate. The learning rate increased linearly over five warm-up epochs and was halved after ten epochs without improvement.

\subsubsection{Gardner parameters and training settings}
\label{sec:closure}

The physics loss requires the two Gardner parameters. The exponent $\beta$ follows from $n$ through \Cref{eq:vgm_k,eq:gardner_se}. The parameter $\alpha_g$ has no exact VGM equivalent, so it was calibrated against numerical trajectories, building on earlier work that relates Gardner and van Genuchten parameters \citep{Rucker2005,Ghezzehei2007}. For each of 512 parameter sets (the first 512 runs of the dataset), the value of $\alpha_g$ that minimised the residual was found, and the following relation was fitted to these optima:

\begin{equation}\label{eq:alphag}
\alpha_g=2.74(n-2)^{0.71}\alpha.
\end{equation}
On a further 512 cases, the residual obtained with \Cref{eq:alphag} was, at the median, 15\% above the per-case optimum (Appendix~\ref{app:alphag}). The final configuration used a batch size of 4096 and a physics weight of $\lambda=0.4$ (\Cref{tab:arch}). The batch size was selected during preliminary model development, and the physics weight was chosen from a comparison of six weights, described in Appendix~\ref{app:sensitivity} and discussed in \Cref{sec:physics_results}. Three networks were trained with different random seeds to measure the variability of training.

\subsection{Predictive irrigation controller}
\label{sec:control}

The controller selects the irrigation rate once a day, as illustrated in \Cref{fig:loop}. At each daily update, it receives the measured saturation profile and the hydraulic parameters of the column, and it considers $N_R=200$ evenly spaced candidate rates between a lower and an upper bound. For each candidate, the surrogate is applied recursively for 24 hourly steps with the rate held constant, which gives the predicted profile at the end of the day. All 200 predictions are computed together as one batch. The controller then selects the rate whose predicted end-of-day profile is closest to the set-point, measured by the mean squared deviation over the column:

\begin{equation}\label{eq:objective}
J(R)=\frac{1}{N_z+1}\sum_{i=0}^{N_z}(\hat S_{e,i}^{24}(R)-S_e^{SP})^2,
\qquad R^\star=\argmin_{R\in\mathcal R}J(R),
\end{equation}
where $\mathcal R$ is the set of candidate rates. This objective rewards a uniform profile at the set-point, which corresponds to homogeneous wetting along the column. The lower bound of $\mathcal R$ is $R_{\mathrm{min}}$, and the upper bound is

\begin{equation}\label{eq:rate_bounds}
 \begin{aligned}
 R_{\mathrm{max}}&=\min\{R_{\mathrm{num}},\;0.30K_s,\;R_{\mathrm{scale,max}}\},\\
 R_{\mathrm{num}}&=\begin{cases}
 0.95R_{\mathrm{stable}},&\text{if the numerical ceiling binds},\\
 0.30K_s,&\text{otherwise}.
 \end{cases}
\end{aligned}
\end{equation}
where $R_{\mathrm{scale,max}}\simeq69.57$~cm/day is the highest rate in the training data. Very high irrigation rates can make the numerical model fail. Before each test, the rate was therefore increased step by step in the simulator to find $R_{\mathrm{stable}}$, the last rate before the first numerical failure. If this limit was below the operational cap of $0.30K_s$, the controller used 95\% of it; otherwise it used the cap. The same bounds were used for every network and controller tested on a given column. The selected rate is applied to the reference simulator for one day, and the resulting profile becomes the starting point for the next decision. This daily restart replaces the predicted state with the measured state, limiting accumulation of state-prediction error; the network weights and the hydraulic parameters are not updated. Each decision therefore requires 24 batched network evaluations and one comparison of 200 candidates.

\subsection{Comparison controllers}
\label{sec:baselines_method}

Four controllers were used for comparison. Three of them are simple rules that use the same known hydraulic parameters as the predictive controller. These rules were built here as reference points rather than taken from the literature, and all their rates were clipped to $[R_{\mathrm{min}},R_{\mathrm{max}}]$. The constant steady-state rule follows from a property of the column model. For a uniform column with free drainage, a profile that sits at the set-point everywhere drains at exactly $K(S_e^{SP})$. Irrigating at $R=K_s(S_e^{SP})^{\delta}$ therefore holds the column at the set-point once it is there. For the reference parameter set, this rate is about 4.0~cm/day. However, it fills a dry column slowly: the initial water deficit of the reference column, $\theta_s(S_e^{SP}-S_{e,\mathrm{init}})H\approx50$~cm of water, would take at least 12 days to fill at this rate. This rule uses no measurements.

The proportional deficit rule adds a correction proportional to the measured water deficit of the column:

\begin{equation}\label{eq:pdrule}
 R_d=K_s\,(S_e^{SP})^{\delta}+K_p\,\theta_sH\,e_d,\qquad e_d=S_e^{SP}-\bar S_{e,d},
\end{equation}
where $\bar S_{e,d}$ is the measured spatial-mean saturation at the start of day $d$, so that $\theta_sH\,e_d$ is the amount of water, in cm, needed to reach the set-point. With $K_p=1$, the rule tries to supply the whole deficit within one day. The PI controller adds integral action, which accumulates the deficit over the previous days:

\begin{equation}\label{eq:pi}
 R_d=K_s\,(S_e^{SP})^{\delta}+\theta_sH\Big(K_p\,e_d+K_i\sum_{j=1}^{d}e_j\Big).
\end{equation}
Integral action can reduce persistent offsets when the target is attainable and actuator limits do not prevent correction. To prevent this term from growing while the rate is held at a bound, an effect known as wind-up, integration was paused whenever the rate was at a bound and the deficit would push it further. The gains, $K_p=0.5$ and $K_i=0.4$, were tuned on 100 validation cases by a grid search over $K_p\in[0.25,1]$ and $K_i\in[0,0.8]$. The selected pair gave the lowest mean of the median tracking errors under correct parameters and under the parameter-error condition at level 0.10 (\Cref{sec:protocol}); the gains were therefore not tuned specifically to minimise overshoot or irrigation.

Finally, a matched-model controller uses the same objective and horizon as the predictive controller, but predicts with the numerical model itself. It isolates the effect of surrogate error, since it uses the same dynamics as the reference process. Because 200 numerical simulations per decision were too slow for this comparison, this controller searched the rate interval by golden-section search instead of the 200-point grid. Differences therefore include the search procedure as well as the prediction model; numerical solver errors are shared with the reference process.

\subsection{Case studies and performance measures}
\label{sec:protocol}

The main case study applied the predictive controller to the 615 evaluation columns with each of the three trained networks, which gave 1845 closed-loop runs of 25 days. Tracking performance is measured by the full-cycle root-mean-square (RMS) deviation from the set-point, expressed as a percentage of the set-point. For the run of case $s$ with network seed $b$,

\begin{equation}\label{eq:tracking_metric}
 E_{s,b}=\frac{100}{S_e^{SP}}\sqrt{\frac{1}{N_h(N_z+1)}
 \sum_{k=1}^{N_h}\sum_{i=0}^{N_z}\big(S_{e,i,s,b}^{k}-S_e^{SP}\big)^2},\qquad N_h=600.
\end{equation}
where $S_{e,i,s,b}^{k}$ is the saturation in the reference simulator at node $i$ and hour $k$. A value of 5\% therefore means that, over the whole cycle and the whole column, saturation differed from 0.70 by about 0.035 in a root-mean-square sense. For each seed, the median of $E_{s,b}$ across cases was taken, and the mean and standard deviation of the three medians are reported. The final saturation, averaged first over the profile and then across cases, is reported in the same way. A run settles on the first day after which its spatial-mean saturation stays within $\pm0.01$ of the set-point.

Four further sets of tests were carried out. First, the physics weight was varied from $\lambda=0$ to 1, and the resulting 18 networks (six weights and three seeds) were compared on the 614 validation cases. Second, each hydraulic parameter was moved beyond its training range to find the limits of reuse. Third, the computing time of the surrogate and of the numerical model was measured on an AMD Ryzen 7 260 CPU and an NVIDIA RTX 5060 Laptop GPU, for one column and for batches of up to 5000 copies of the reference column. Fourth, the predictive controller was compared with the controllers of \Cref{sec:baselines_method} and tested with imperfect information. Three imperfect-information conditions were examined: Gaussian noise with a standard deviation of 0.02 added to every node of the measured profile; only five equally spaced sensors, with noise of 0.01 and linear interpolation between them; and errors in the hydraulic parameters. For parameter errors, each of $K_s$, $\alpha$, $n$ and $\theta_s$ supplied to the controller was multiplied by an independent log-normal factor, at levels of 0.10 and 0.20, while the reference simulator retained the true values; these results are reported in Appendix~\ref{app:stress}. The simple rules were run on all 615 evaluation cases. Because of computing limits, the matched-model controller and the tests with imperfect information used a random subset of 205 evaluation cases and the seed-42 network. For these comparisons, the closed loop was re-implemented on a CPU and checked against the original nominal irrigation decisions. Paired comparisons report the share of non-tied cases in which one controller was better; because cases and networks are shared, these shares are descriptive and no binomial confidence intervals are given.

\section{Results and discussion}
\label{sec:results}

\subsection{Saturation control across hydraulic conditions}
\label{sec:generalisation}
\label{sec:closedloop}

\Cref{tab:closedloop} summarises the performance of the predictive controller on the 615 evaluation columns. Without any retraining, the same surrogate was used for all evaluated columns. The median full-cycle tracking deviation was $4.9\pm0.1$\% of the set-point, and the final saturation averaged $0.697\pm0.003$ (median $0.699\pm0.002$). The deviation includes the initial wetting period, when the column is still far from the set-point; almost all of it arose in the first three days (97\% of the end-of-day squared deviation, median over runs).

\begin{table}[H]
  \centering
  \caption{Closed-loop results on the 615 evaluation cases. The RMS deviation is the median across cases and the final saturation the mean across cases, both given as mean $\pm$ standard deviation over three seeds. The settling time refers to the example in \Cref{fig:panel} (spatial-mean saturation within $\pm0.01$ of the set-point).}
  \label{tab:closedloop}
  \small
  \setlength{\tabcolsep}{6pt}
  \renewcommand{\arraystretch}{1.15}
  \begin{tabular}{@{} l l l l @{}}
    \toprule
    \textbf{Configuration} & \textbf{Value} & \textbf{Performance measure} & \textbf{Value} \\
    \midrule
    evaluation cases / closed-loop runs & 615 / 1845 & final mean $S_e$ & $0.697 \pm 0.003$ \\
    leach cycle / control step & 25 d / 24 h & RMS deviation [\% of $S_e^{SP}$] & $4.9 \pm 0.1$ \\
    candidate rates / set-point & 200 / 0.70 & settling time of example [d] & 4 \\
    \bottomrule
  \end{tabular}
\end{table}

\begin{figure}[!htbp]
  \centering
  \includegraphics[width=1\linewidth,height=0.9\textheight,keepaspectratio]{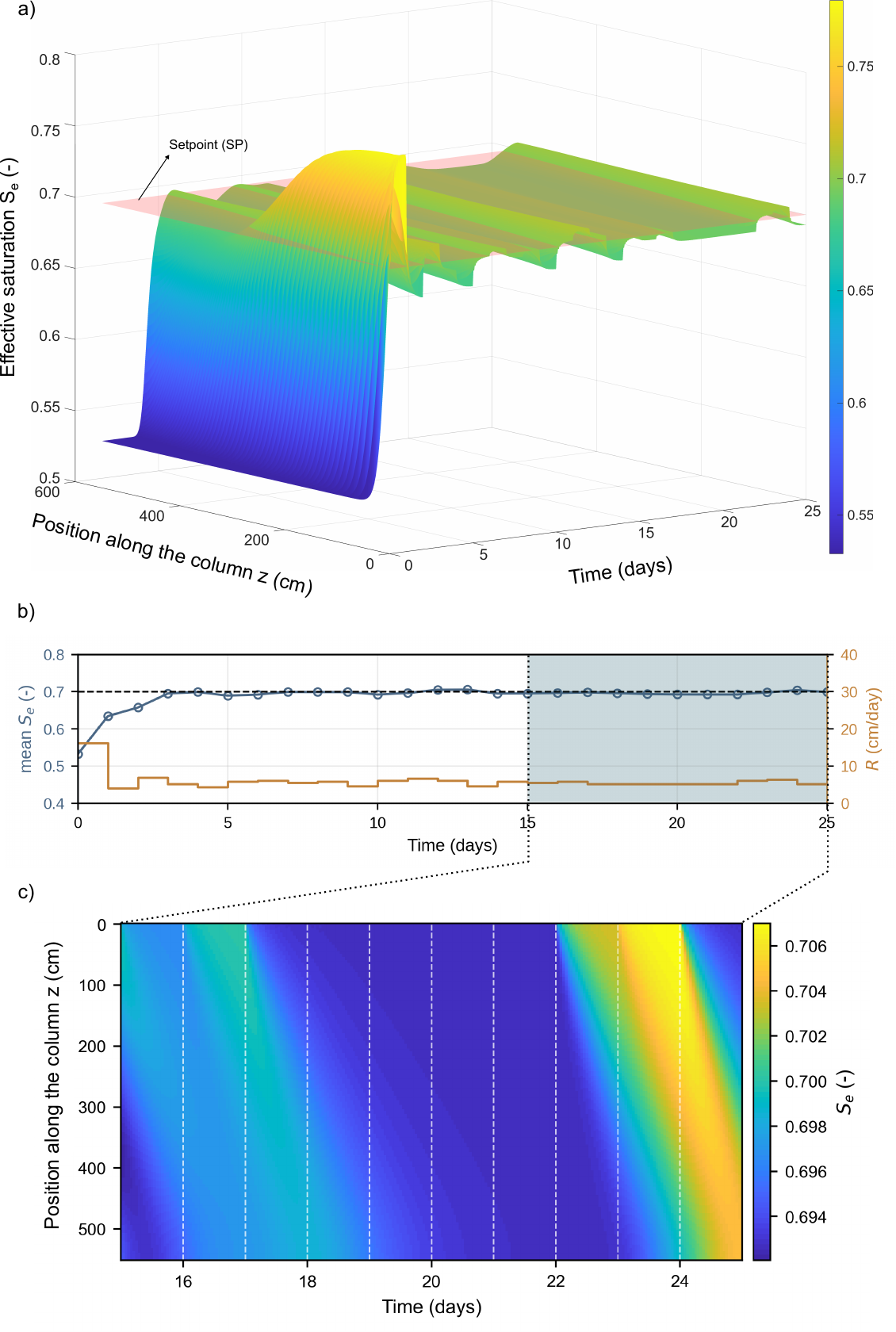}
  \caption{Representative closed-loop run for evaluation case 2917 (seed 42; $K_s = 192.6$~cm/day, $\alpha = 0.0333$~cm$^{-1}$, $n = 2.293$, $\theta_s = 0.287$, $S_{e,\mathrm{init}} = 0.533$). (a) Saturation profile along the column over the leach cycle, set-point plane marked at 0.70. (b) Spatial-mean saturation and applied irrigation rate from the initial condition $S_{e,\mathrm{init}}=0.533$ at day~0; the shaded window highlights the period in (c). (c) Heat map of the last ten days on a narrowed colour scale, with 24-hour control steps separated by dashed lines.}
  \label{fig:panel}
\end{figure}

A representative evaluation case is shown in \Cref{fig:panel}. The column wets from the top downwards during the first days, and its spatial-mean saturation stays within $\pm0.01$ of the set-point from day four onwards. After that, the daily decisions hold a nearly uniform profile close to 0.70 (\Cref{fig:panel}c). Across all 1845 runs, the median settling time was two days. This behaviour is consistent with the single-column NMPC of \citet{Olivares2025}, which also held a uniform profile at the set-point once moisture measurements were used for feedback. Here, however, one model served all 615 columns without any column-specific fitting.

Performance was not uniform across the hydraulic range. About 1.5\% of runs (28 of 1845) had a deviation above 15\%, and all of them came from columns with a low $n$ (at most 2.38) and a dry start ($S_{e,\mathrm{init}}<0.47$). A low $n$ means that the conductivity falls steeply as the bed dries (\Cref{eq:vgm_k}), so a dry column of this type wets slowly even at the maximum allowed rate. In 2.1--3.7\% of evaluation cases, depending on the seed, the reference simulator also had at least one hour with a bound violation; these runs were kept in all summaries. Limited wetting capacity, surrogate error and numerical error may all contribute to these difficult cases.

\subsection{Comparison with other controllers}
\label{sec:baselines}

\Cref{tab:baselines} compares the predictive controller with the controllers of \Cref{sec:baselines_method}, with known parameters and full measurements. The predictive controller and the tuned PI controller tracked the set-point equally well. Both had a median deviation of 4.9\%; the predictive controller had the lower deviation in 57\% of paired cases, and the median difference was only 0.06 percentage points. Both were better than the proportional deficit rule (5.4\%) and the constant rule (5.7\%), which the predictive controller outperformed in 69\% and 91\% of paired cases, respectively.

\begin{table}[H]
  \centering
  \caption{Comparison with simple controllers and a matched-model controller, with known parameters and the full noise-free profile. $E$ is the full-cycle RMS deviation (\Cref{eq:tracking_metric}); $|\Delta S|$ is the median distance of the final spatial-mean saturation from the set-point; bound violations and upper-bound use give the share of runs with at least one event; water is the mean cumulative applied irrigation over 25 days. The first row gives the mean over three seeds (ranges for the event shares).}
  \label{tab:baselines}
  \footnotesize
  \setlength{\tabcolsep}{4pt}
  \renewcommand{\arraystretch}{1.2}
  \begin{tabular}{@{} l c c c c c c c c @{}}
    \toprule
    \textbf{Controller} & \textbf{Cases} & \textbf{median $E$} & \textbf{90th pct $E$} & \textbf{$E>15$\%} & \textbf{$|\Delta S|$} & \textbf{bound viol.} & \textbf{at $R_{\mathrm{max}}$} & \textbf{water} \\
     & & [\%] & [\%] & [\% runs] & [--] & [\% runs] & [\% runs] & [cm] \\
    \midrule
    PGML predictive controller (this work) & 615 & 4.90 & 7.41 & 1.5 & 0.002 & 2.1--3.7 & 4.9--10.1 & 449 \\
    constant steady-state rate & 615 & 5.75 & 11.85 & 5.7 & 0.000 & 0.0 & 0.0 & 436 \\
    proportional deficit rule & 615 & 5.38 & 7.05 & 0.5 & 0.000 & 11.2 & 46.2 & 462 \\
    PI controller (tuned) & 615 & 4.87 & 7.22 & 0.3 & 0.000 & 6.3 & 9.1 & 469 \\
    \midrule
    PGML predictive controller (seed 42) & 205 & 4.59 & 7.96 & 2.0 & 0.002 & 2.4 & 7.3 & 437 \\
    matched-model predictive controller & 205 & 4.58 & 7.02 & 0.0 & 0.000 & 5.4 & 14.6 & 448 \\
    \bottomrule
  \end{tabular}
\end{table}

\begin{figure}[!htbp]
  \centering
  \includegraphics[width=0.95\linewidth]{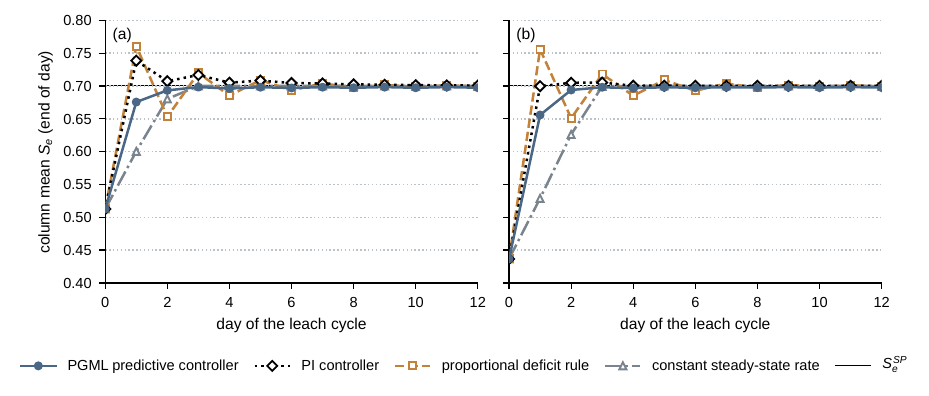}
  \caption{Median column-mean saturation at the end of each day for the predictive controller, the PI controller, the proportional deficit rule and the constant steady-state rate, with known parameters: (a) all 615 evaluation cases; (b) the 201 cases with a dry start ($S_{e,\mathrm{init}}<0.47$). Day 0 is the initial state. Days 13--25 are not shown; the plotted median trajectories stay close to the set-point.}
  \label{fig:approach}
\end{figure}

The controllers differed more in how they reached the set-point, as \Cref{fig:approach} shows. In the median, the predictive controller brought the column to within 0.01 of the set-point by day two and kept it there. The PI median trajectory wetted the columns faster initially, overshot to 0.74 on the first day and remained within $\pm0.01$ only after about five days. The proportional deficit rule had a median trajectory that overshot to 0.76 and then oscillated around the set-point for about a week. The constant rule was the slowest, especially for dry starts: for cases with $S_{e,\mathrm{init}}<0.47$, its median deviation was 8.8\%, compared with 6.7\% for the predictive controller.

The PGML controller had at least one numerical bound violation in 2--4\% of runs, compared with 6.3\% for PI and 11\% for the proportional deficit rule. The share of runs reaching the upper irrigation bound was 5--10\%, similar to the PI value of 9.1\% and well below the proportional rule's 46\%. Mean cumulative applied irrigation was 449~cm for PGML and 469~cm for PI, approximately 4\% lower; this is gross applied irrigation in the simulations, not net water consumption in a recirculating circuit. The matched-model controller gave a similar median deviation on the 205-case subset (4.58\% versus 4.59\%), but a lower 90th-percentile error (7.02\% versus 7.96\%) and no runs above 15\%, so surrogate error matters mainly in the difficult cases. Overall, for a fixed target and known hydraulic parameters, a tuned PI controller is a strong alternative: the two controllers track equally well, and the advantages of the predictive controller lie in a smoother approach to the target, slightly lower irrigation and, above all, the ability to evaluate many candidate actions per column at low cost (\Cref{sec:cost}).

\subsection{Sensitivity to measurement and parameter errors}
\label{sec:robust}

With synthetic measurement errors, the median PGML tracking deviation changed from 4.59\% to 4.62\% with node-wise noise of standard deviation 0.02, and to 4.63\% with five equally spaced sensors and noise of 0.01. PI deviations were 4.66\%, 4.67\% and 4.70\%, respectively (\Cref{tab:measurement}). The PGML controller had lower deviation than the proportional rule in 69\% and 73\% of paired cases under the two perturbed conditions, and than PI in 55--60\%. Both controllers were therefore insensitive to these synthetic perturbations, although correlated sensor errors and a real ERT measurement pipeline were not examined. Errors in the hydraulic parameters had a larger effect (Appendix~\ref{app:stress}): at level 0.10 the median PGML deviation rose to 8.0\%, whereas PI remained at 4.8\%. The predictive controller uses fixed parameters and has no integral correction, so a parameter mismatch can produce a persistent offset; online parameter estimation or offset-free formulations are natural extensions.

\begin{table}[H]
\centering
\caption{Measurement sensitivity on 205 evaluation cases with the seed-42 network. Entries are median full-cycle tracking deviation, in percent of the set-point. Hydraulic parameters are correct in all rows.}
\label{tab:measurement}
\small
\begin{tabular}{@{}lccc@{}}
\toprule
Condition & PGML & PI & Proportional deficit \\
\midrule
Full profile, no noise & 4.59 & 4.66 & 5.26 \\
Full profile, noise sd 0.02 & 4.62 & 4.67 & 5.26 \\
Five sensors, noise sd 0.01 & 4.63 & 4.70 & 5.26 \\
\bottomrule
\end{tabular}
\end{table}

\subsection{Effect of physics guidance}
\label{sec:physics_results}

The physics weight comparison asks whether the mass-balance term adds anything beyond the data term. Six weights, from purely data-driven ($\lambda=0$) to purely physics-based ($\lambda=1$) training, were trained with three seeds and tested on the same validation cases. For two weights, the win fraction is the share of non-tied paired runs in which one weight performs better than the other (Appendix~\ref{app:lambda}). The physics term had three effects. First, it improved long-horizon prediction. In open-loop predictions over the full 600 hours, without any correction from measurements, the mean prediction error $\bar\varepsilon$ (\Cref{eq:eps}) was $1.26\times10^{-2}$ for data-only training and $0.83$--$0.92\times10^{-2}$ for $\lambda=0.2$--0.6, about 30\% lower; at $\lambda=0.4$, the hybrid network had the lower error in approximately two-thirds of paired cases. Second, it reduced the number of runs with upper-bound activity and numerical bound violations. Compared with data-only training, $\lambda=0.4$ won 94\% of non-tied comparisons on numerical bound violations (107 non-tied pairs) and 89\% on use of the upper irrigation bound (274 pairs). Runs with at least one day at the upper bound fell from 331 to 116 of 1842, and runs with at least one bound violation fell from 141 to 45 (\Cref{fig:lambda_rmax}). Third, it slightly worsened closed-loop tracking. The median tracking deviation on the validation cases was 4.56\% for data-only training and 4.65\% for $\lambda=0.4$, and data-only training had the lower deviation in 66\% of paired cases (1839 pairs), although the difference is small.

Better long-horizon prediction did not translate into lower tracking error. A plausible explanation is that the daily measured-state reset limits error accumulation, so the 600-hour open-loop accuracy matters less than the ranking of candidates over 24 hours. Physics-only training ($\lambda=1$) had the largest open-loop error ($3.06\times10^{-2}$) and median tracking deviation (4.98\%), consistent with the Gardner constraint only approximating the VGM dynamics. Moderate physics guidance therefore produced a trade-off between open-loop prediction, tracking and numerical diagnostics.

\subsection{Operating range beyond the training domain}
\label{sec:extrap}

\begin{figure}[!htbp]
  \centering
  \includegraphics[width=0.9\linewidth]{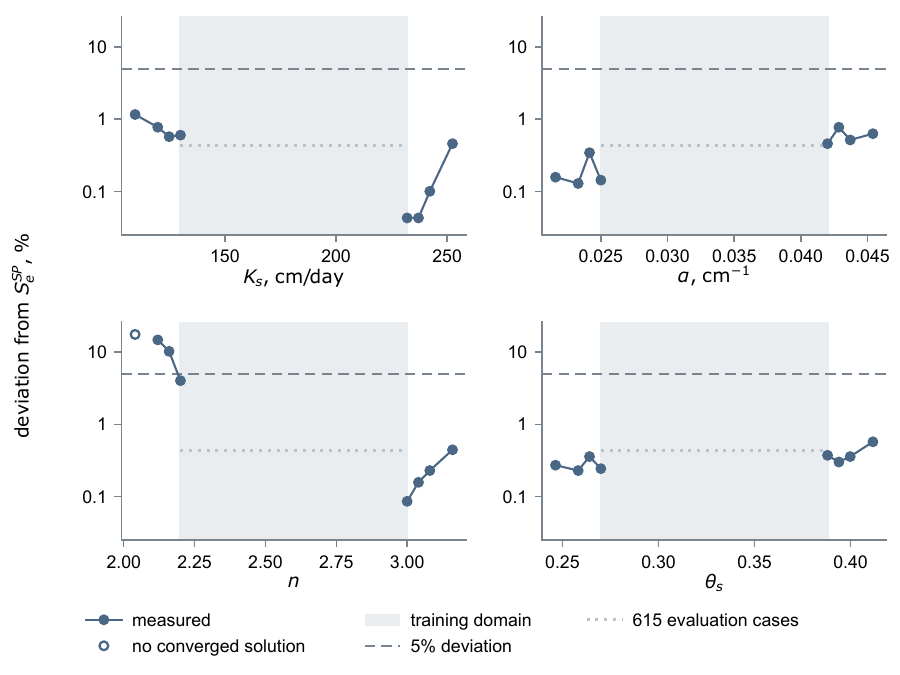}
  \caption{Final deviation from set-point against each parameter value. Shading marks the training range; the dashed line marks 5\%. Filled markers (labelled `measured' in the legend) are closed-loop simulation results. The dotted line shows the evaluation-set final deviation, not the full-cycle RMS. The open marker shows a numerical failure; its height is for display only.}
  \label{fig:extrap_single}
\end{figure}

To test how far the trained surrogate can be reused outside its training range, each hydraulic parameter was moved past its lower or upper bound. For a parameter $p\in[p_L,p_U]$, the tested values were $p_L-\gamma(p_U-p_L)$ and $p_U+\gamma(p_U-p_L)$, with an excursion fraction $\gamma\in\{0,0.05,0.10,0.20\}$. Excursions are therefore expressed as a percentage of the width of the training range. In each test, the other parameters and the initial saturation came from the same 256-case Sobol sample. The aggregate terminal deviation remained below 1.2\% in seven of the eight parameter directions, even at the largest excursion (\Cref{tab:extrap_single,fig:extrap_single}). The exception was the lower bound of $n$. As $n$ fell from 2.20 to 2.16 and 2.12, the final deviation rose from 4.01\% to 10.21\% and 14.73\%, and the simulation at $n=2.04$ failed. Reuse outside the training range therefore depends on the direction, and low $n$ is the weak point. This limit is relevant in practice, because the reference value of $n=2.267$ \citep{Cariaga2015} lies only 0.067 above the lower bound of the training range.

\begin{table}[H]
  \centering
  \caption{Final deviation from set-point (\%) when one parameter is moved beyond its training range. Column headings give the excursion as a percentage of the training-range width. $\dagger$ marks a case where the numerical solution failed in our setup; it does not prove that the case is physically infeasible.}
  \label{tab:extrap_single}
  \footnotesize
  \setlength{\tabcolsep}{9pt}
  \renewcommand{\arraystretch}{1.25}
  \begin{tabular}{@{} l l c c c c @{}}
    \toprule
    & & \multicolumn{4}{c}{excursion beyond the bound} \\
    \cmidrule(l){3-6}
    \textbf{Property} & \textbf{Bound} & \textbf{0\%} & \textbf{5\%} & \textbf{10\%} & \textbf{20\%} \\
    \midrule
    $K_s$, cm/day        & $\leq 130$    & 0.60 & 0.57 & 0.77 & 1.16 \\
                         & $\geq 232$    & 0.04 & 0.04 & 0.10 & 0.46 \\
    $\alpha$, cm$^{-1}$  & $\leq 0.025$  & 0.14 & 0.34 & 0.13 & 0.16 \\
                         & $\geq 0.042$  & 0.46 & 0.77 & 0.51 & 0.63 \\
    $n$                  & $\leq 2.20$   & 4.01 & 10.21 & 14.73 & $\dagger$ \\
                         & $\geq 3.00$   & 0.09 & 0.16 & 0.23 & 0.44 \\
    $\theta_s$           & $\leq 0.270$  & 0.24 & 0.36 & 0.23 & 0.27 \\
                         & $\geq 0.388$  & 0.37 & 0.30 & 0.36 & 0.57 \\
    \bottomrule
  \end{tabular}
\end{table}

The VGM conductivity exponent offers a physical explanation for sensitivity near the lower bound of $n$. As $n\to2^+$, $m=1-2/n\to0$ and $\delta\to\infty$, so the conductivity at any fixed $0<S_e<1$ falls sharply. All tested values still satisfied $n>2$, but the column became slow to wet and more difficult to simulate and predict. Moving all four properties at once by 10\% of their range widths gave the same picture (Appendix~\ref{app:extrap}). The pooled terminal mean for the eight combinations with $n$ above its upper bound was near the target, whereas the group with $n$ below its lower bound gave a pooled final deviation of 14.7\%. Near $n=2$, target attainability must also be checked against the irrigation bounds through $R_{\mathrm{eq}}=K_s(S_e^{SP})^\delta$. Extending the training range towards lower $n$ is therefore a priority for future versions of the surrogate.

\subsection{Computational cost}
\label{sec:cost}

The cost of a matched recursive-prediction workload is shown in \Cref{tab:speed}. For 200 candidate rates over 24 hourly steps, the surrogate took $15.9\pm0.8$~ms on the GPU and $34.9\pm8.0$~ms on the CPU, whereas the numerical model took $1150\pm20$~ms and $2560\pm35$~ms for the same task. For this matched 200-candidate workload, the surrogate was about 70 times faster on both devices, a 98.6\% reduction in prediction time. This figure excludes training and parameter identification. For comparison, \citet{Olivares2025} reported mean identification times of 1.398--3.454~s and mean optimisation times of 0.063--0.095~s per decision on their own hardware. Their calculations include online model updating and solve a different control problem, so the numbers are not directly comparable, but they confirm that model-based control of a single column fits easily within a daily interval. The same is true of the numerical model here: about one second per decision is fast enough for one column.

\begin{table}[H]
  \centering
  \caption{Recursive-prediction time for one column (mean $\pm$ standard deviation, ms) for 200 candidates over a 24-hour horizon, from 15 repetitions after 10 warm-up runs.}
  \label{tab:speed}
  \small
  \setlength{\tabcolsep}{10pt}
  \renewcommand{\arraystretch}{1.15}
  \begin{tabular}{@{} l c c @{}}
    \toprule
    \textbf{Prediction model} & \textbf{CPU [ms]} & \textbf{GPU [ms]} \\
    \midrule
    PGML (this work)     & $34.9 \pm 8.0$ & $15.9 \pm 0.8$ \\
    Numerical solver     & $2560 \pm 35$  & $1150 \pm 20$  \\
    \bottomrule
  \end{tabular}
\end{table}

\begin{figure}[!htbp]
  \centering
  \includegraphics[width=0.85\linewidth]{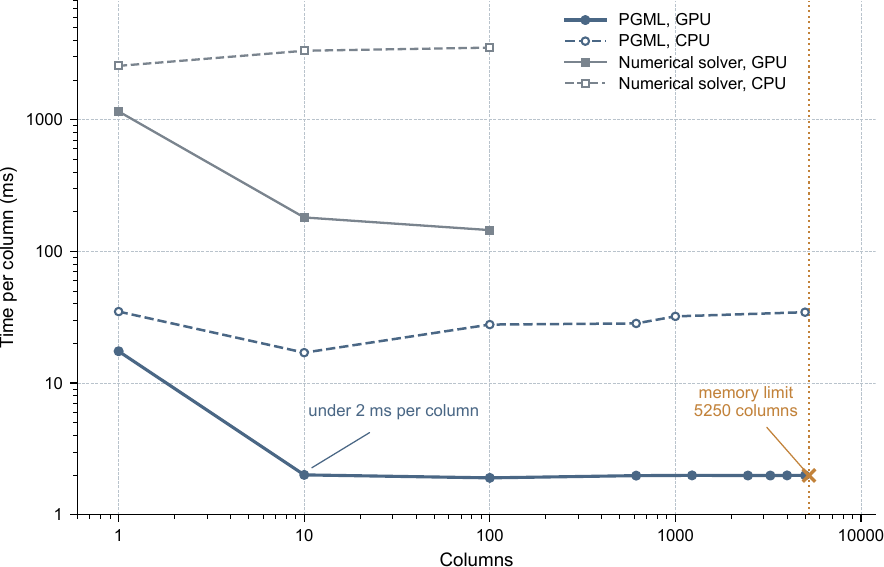}
  \caption{Average prediction cost per column against the number of columns in a batch. Total batch time equals the plotted cost multiplied by the number of columns. Points average 15 repetitions after 10 warm-up runs; the cross marks the batch size that ran out of GPU memory.}
  \label{fig:speed_scaling}
\end{figure}

The speed of the surrogate becomes useful when many columns must be controlled. Batching lowered the GPU cost further, and the average cost stayed below 2~ms per column up to 5000 columns (\Cref{fig:speed_scaling}), so a batch of 5000 columns took less than 10~s in total. Sequentially repeating the single-column numerical workload 5000 times would take approximately 1.6~hours. This is a sequential extrapolation; a numerical solver could also be parallelised. On the CPU, batching helped little, with costs of 27.8--32.1~ms per column at large batch sizes. The achievable scale depends on the horizon, the number of candidates, the hardware and the memory, as the memory limit in \Cref{fig:speed_scaling} shows. These timings use copies of one parameter set and benchmark batched prediction rather than coordinated closed-loop control of many columns.

\subsection{Practical implications and limitations}
\label{sec:limitations}

The main contribution is a single parameter-conditioned surrogate that serves columns with different hydraulic properties without retraining, combined with a description of the trade-offs between prediction accuracy, tracking and numerical behaviour. The surrogate does not perform parameter identification, so the hydraulic properties and the saturation state must still be supplied, for example from ERT measurements and an estimation step such as that of \citet{Olivares2025}. With correct parameters, nominal tracking was equal to that of a tuned PI controller, with a smoother approach and lower applied irrigation; with erroneous parameters, PI was more robust. Moderate physics guidance improved long-horizon prediction and reduced constraint activity while slightly worsening tracking relative to data-only training.

\section{Conclusions}
\label{sec:conclusions}

This study developed a parameter-conditioned PGML surrogate for simulated heap-leaching hydraulics and evaluated it in a daily predictive controller. Trained on 2867 trajectories from a 4096-case dataset, one network served 615 evaluation columns with different hydraulic properties and achieved a median tracking deviation of $4.9\pm0.1$\% of a 0.70 saturation target. This was equal to a tuned PI controller, while the PGML median trajectory approached the target without overshoot and applied approximately 4\% less irrigation. Tracking was insensitive to the synthetic measurement noise and sparse sensors tested, but deteriorated when the controller was given erroneous hydraulic parameters, where PI was more robust.

Moderate physics guidance reduced long-horizon prediction error by approximately 30\% relative to data-only training and reduced the number of runs reaching the upper irrigation bound from 331 to 116 of 1842, at the cost of a slight loss in closed-loop tracking. Reuse beyond the training range was robust in seven of eight parameter directions but failed for low $n$. The surrogate was approximately 70 times faster than the numerical solver for a matched 200-candidate, 24-hour workload, and batched GPU evaluation cost less than 2~ms per column for 5000 columns. These results support the development of the surrogate towards heap-scale control, with online parameter estimation, an extended training range, independent numerical verification and experimental validation as the next steps.

\section*{CRediT authorship contribution statement}
\textbf{Sultan Alhamdan}: Methodology, Software, Formal analysis, Investigation, Data curation, Writing -- original draft, Visualization. \textbf{Daniel Navia}: Conceptualization, Methodology, Writing -- review \& editing. \textbf{Paulina Quintanilla}: Conceptualization, Methodology, Formal analysis, Supervision, Writing -- original draft, Writing -- review \& editing, Project administration.

\section*{Data and code availability}
The public repository provides the reference simulator, data-generation pipeline, training code and PGML closed-loop controller at \url{https://github.com/Sultan-Hamdan/PGML-Copper-Heap-Leaching}.

\section*{Declaration of competing interest}
The authors declare that they have no known competing financial interests or personal relationships that could have appeared to influence the work reported in this paper.

\bibliography{references}

\newpage
\appendix
\makeatletter
\@namedef{appx@pre@section}{Appendix~}
\renewcommand{\@seccntformat}[1]{%
  \@nameuse{appx@pre@#1}\csname the#1\endcsname\hspace{1em}%
}
\makeatother
\counterwithin{table}{section}
\counterwithin{figure}{section}
\counterwithin{equation}{section}
\renewcommand{\thetable}{\Alph{section}\arabic{table}}
\renewcommand{\thefigure}{\Alph{section}\arabic{figure}}
\renewcommand{\theequation}{\Alph{section}\arabic{equation}}

\section{Discrete physics residual}
\label{app:residual}

The physics residual is evaluated in moisture content, with $\theta_i^k=\theta_sS_{e,i}^k$ and $\hat\theta_i^{k+1}=\theta_s\hat S_{e,i}^{k+1}$ for $i=0,\ldots,N_z$. Constitutive factors use $\widetilde S_e=\min(1,\max(10^{-9},S_e))$ in \Cref{eq:gardner_se}; the linear moisture terms retain their unmodified values. At either time level $\ell\in\{k,k+1\}$, arithmetic face averages give the downward interior flux

\begin{equation}\label{eq:face_flux}
 q_{i+1/2}^{\ell}=\frac{K_{G,i}^{\ell}+K_{G,i+1}^{\ell}}{2}
 -\frac{D_{G,i}^{\ell}+D_{G,i+1}^{\ell}}{2}
   \frac{\theta_{i+1}^{\ell}-\theta_i^{\ell}}{\Delta z}.
\end{equation}
The boundary fluxes are $q_{-1/2}^{\ell}=R^k$ at the surface and $q_{N_z+1/2}^{\ell}=K_{G,N_z}^{\ell}$ at the base. The resulting Crank--Nicolson moisture-increment residual is

\begin{equation}\label{eq:physics_residual}
 r_i=\hat\theta_i^{k+1}-\theta_i^k
 -\frac{\Delta t}{2\Delta z}
 \left(q_{i-1/2}^{k}-q_{i+1/2}^{k}
       +q_{i-1/2}^{k+1}-q_{i+1/2}^{k+1}\right),
\end{equation}
Future fluxes are evaluated at the surrogate prediction. For a batch of $B$ transitions indexed by $b$, the raw losses are

\begin{equation}\label{eq:raw_losses}
 \mathcal L_p=\frac{1}{B(N_z+1)}\sum_{b=1}^{B}\sum_{i=0}^{N_z}r_{i,b}^2,
 \qquad
 \mathcal L_d=\frac{1}{B(N_z+1)}\sum_{b=1}^{B}\sum_{i=0}^{N_z}
 (\hat S_{e,i,b}^{k+1}-S_{e,i,b}^{k+1})^2.
\end{equation}
The discretisation uses $\Delta t=1/24$~day and $\Delta z=2.5$~cm. Flux is in cm/day, and $r$ is in units of volumetric moisture content. The residual gives each boundary node a full cell width $\Delta z$. This differs from a vertex-centred half-cell convention and from the ghost-node treatment in the VGM simulator, so any conservation or grid-convergence check must treat each discrete operator separately.

\section{Gardner diffusivity parameter calibration}
\label{app:alphag}

\begin{figure}[H]
    \centering
    \includegraphics[width=0.65\linewidth]{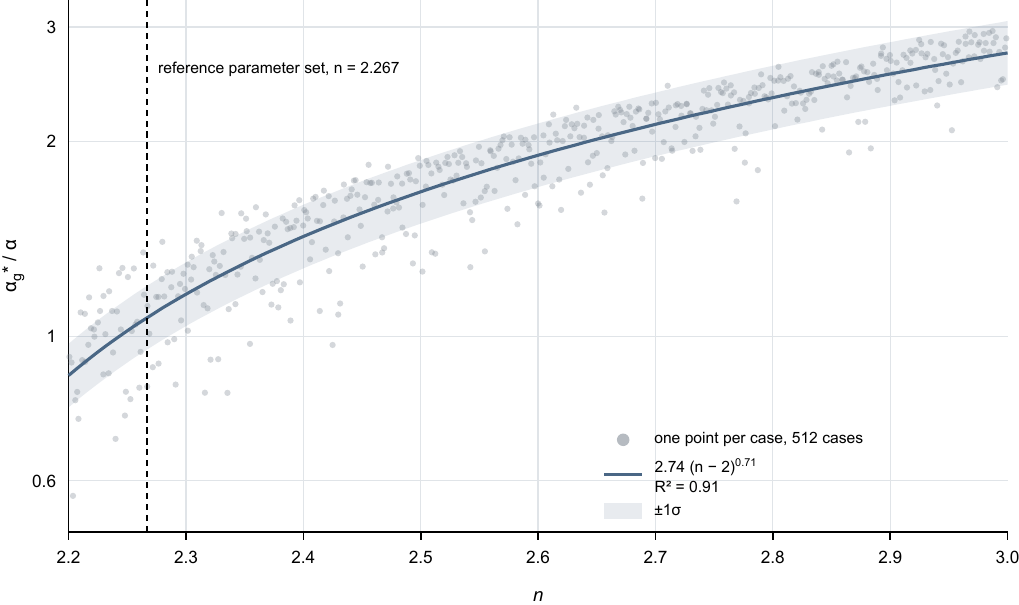}
    \caption{Per-case optima $\alpha_g^{*}$, divided by $\alpha$, plotted against $n$ for the 512 cases of the search, with \Cref{eq:alphag} shown as the solid line. The band spans one standard deviation of the optima about the line; the dashed line marks the reference parameter set at $n=2.267$.}
    \label{fig:alphag_fit}
\end{figure}

With $\beta$ fixed, $\alpha_g$ sets the Gardner diffusivity, $D_G(S_e)=\beta K_sS_e^{\beta-1}/(\alpha_g\theta_s)$. Because $\alpha_g$ has units of inverse length, we scale it by $\alpha$, and we capture its dependence on the modified closure through $n-2$:
\begin{equation}\label{eq:alphag_form}
    \alpha_g = A\,(n-2)^{p}\,\alpha,
\end{equation}
The fit uses the residual-minimising value of $\alpha_g$ for each of 512 Sobol-sampled parameter sets within \Cref{tab:envelope}. For each case, a log-spaced search over four orders of magnitude is refined by fitting a quadratic through the three lowest-residual points in log space. Unlike direct matching of constitutive curves \citep{Ghezzehei2007,Rucker2005}, this calibration evaluates the discrete residual on the profile that the simulator produces one step later. Taking logarithms gives
\begin{equation}\label{eq:alphag_log}
    \log\!\left(\frac{\alpha_g}{\alpha}\right) = \log A + p\,\log(n-2),
\end{equation}
Ordinary least squares gives $A=2.74$ and $p=0.71$ (\Cref{eq:alphag,fig:alphag_fit}). On a further 512 cases, the residual with the fitted relation is, at the median, 15\% above the per-case optimum. This measures the cost of using one shared relation instead of per-case values, within the sampled range and for this discrete residual.

\section{Hyperparameter sensitivity study}
\label{app:sensitivity}

\subsection{Training batch size}
\label{app:batch}

The batch size and physics weight in \Cref{tab:arch} were chosen as follows. Batch size 4096 was selected during preliminary model development, based on prediction results on the 615 evaluation cases and on training time. That comparison was made with an earlier version of the training code and is not reported quantitatively here, and no claim of an optimal batch size is made. For the physics-weight comparison below, the open-loop metric is the time average of the spatial RMSE averaged across cases:
For the physics-weight comparison below, the open-loop metric is the time average of the spatial RMSE averaged across cases:
\begin{equation}\label{eq:eps}
\varepsilon_s(k)=\sqrt{\frac{1}{N_z+1}\sum_{i=0}^{N_z}(\hat S_{e,i,s}^{k}-S_{e,i,s}^{k})^2},\qquad
\bar\varepsilon=\frac{1}{N_h}\sum_{k=1}^{N_h}\frac{1}{N_s}\sum_{s=1}^{N_s}\varepsilon_s(k).
\end{equation}
Here $N_s=614$ for the validation comparison of physics weights and $N_h=600$ for the reported cycle. The quantity $100\bar\varepsilon$ is in percentage points of saturation, unlike the tracking error $E$, which is expressed relative to the 0.70 target.

\subsection{Physics weight}
\label{app:lambda}

Six physics weights, $\lambda\in\{0,0.2,0.4,0.6,0.8,1\}$, were each trained with three seeds, giving 18 networks. All other settings follow \Cref{tab:arch}. Each network was tested on the same 614 validation cases. Comparisons pair the same case and seed, giving 1842 pairs before ties are removed. For a metric where lower is better, let $\mathcal I_{AB}=\{i:m_A^i\ne m_B^i\}$ be the set of non-tied pairs. The win fraction of surrogate $A$ against $B$ is

\begin{equation}\label{eq:psup}
 \widehat p_{AB}=\frac{1}{|\mathcal I_{AB}|}
 \sum_{i\in\mathcal I_{AB}}\mathbf 1\{m_A^i<m_B^i\}.
\end{equation}
If all pairs are tied, the fraction is undefined. The 0.75 threshold is descriptive, following the use of probability-of-superiority comparisons in machine-learning benchmarks \citep{Bouthillier2021}. Tracking and open-loop errors are almost never tied (at least 1839 non-tied pairs out of 1842). The event counts are often tied, mostly because neither run had an event: the comparisons on bound violations rest on 54--118 non-tied pairs, and those on upper-bound use on 129--374. Because cases and networks are shared, the pooled pairs are not independent trials, so fractions are reported descriptively without confidence intervals.

\subsubsection{Open-loop and closed-loop tracking}

The open-loop comparison measures $\bar\varepsilon$ over the full cycle with no state correction. Every setting with a data term passes the 0.75 threshold against physics-only training (\Cref{tab:lambda_openloop}). The hybrid settings $\lambda=0.2$--0.6 also beat data-only training in 68--74\% of pairs, and their mean errors ($0.83$--$0.92\times10^{-2}$) are about 30\% below the data-only value ($1.26\times10^{-2}$); for $\lambda=0.8$ and 1 the mean errors are $1.91$ and $3.06\times10^{-2}$. The closed-loop comparison measures the accumulated deviation from the set-point with daily state feedback, and no pair passes the threshold on this measure (\Cref{tab:lambda_tracking}); data-only training is slightly ahead of every other setting. Better open-loop prediction therefore does not automatically give better tracking when the state is corrected every day.

\begin{table}[H]
  \centering
  \caption{Physics-weight comparison, open loop. Each entry is the win fraction (\Cref{eq:psup}) of $\lambda_A$ against $\lambda_B$; all 1842 paired runs are non-tied. Entries $\geq0.75$ are in bold (descriptive threshold).}
  \label{tab:lambda_openloop}
  \footnotesize
  \setlength{\tabcolsep}{7pt}
  \renewcommand{\arraystretch}{1.25}
  \begin{tabular}{@{} l c c c c c c @{}}
    & \multicolumn{6}{c}{$\lambda_B$} \\
    \cmidrule(l){2-7}
    $\lambda_A$ & \textbf{0.00} & \textbf{0.20} & \textbf{0.40} & \textbf{0.60} & \textbf{0.80} & \textbf{1.00} \\
    \midrule
    \textbf{0.00} & \sub{--}   & \sub{0.26} & \sub{0.32} & \sub{0.33} & \sub{0.46} & \textbf{0.80} \\
    \textbf{0.20} & \sub{0.74} & \sub{--}   & \sub{0.56} & \sub{0.50} & \sub{0.71} & \textbf{0.93} \\
    \textbf{0.40} & \sub{0.69} & \sub{0.44} & \sub{--}   & \sub{0.47} & \sub{0.63} & \textbf{0.91} \\
    \textbf{0.60} & \sub{0.68} & \sub{0.50} & \sub{0.53} & \sub{--}   & \sub{0.66} & \textbf{0.90} \\
    \textbf{0.80} & \sub{0.55} & \sub{0.29} & \sub{0.37} & \sub{0.34} & \sub{--}   & \textbf{0.84} \\
    \textbf{1.00} & \sub{0.20} & \sub{0.07} & \sub{0.09} & \sub{0.10} & \sub{0.16} & \sub{--} \\
    \bottomrule
  \end{tabular}
\end{table}

\begin{table}[H]
  \centering
  \caption{Physics-weight comparison, closed loop (deviation from $S_e^{SP}$ accumulated over the cycle). No entry reaches the 0.75 threshold. Ties are excluded.}
  \label{tab:lambda_tracking}
  \footnotesize
  \setlength{\tabcolsep}{7pt}
  \renewcommand{\arraystretch}{1.25}
  \begin{tabular}{@{} l c c c c c c @{}}
    & \multicolumn{6}{c}{$\lambda_B$} \\
    \cmidrule(l){2-7}
    $\lambda_A$ & \textbf{0.00} & \textbf{0.20} & \textbf{0.40} & \textbf{0.60} & \textbf{0.80} & \textbf{1.00} \\
    \midrule
    \textbf{0.00} & \sub{--}   & \sub{0.69} & \sub{0.66} & \sub{0.64} & \sub{0.67} & \sub{0.72} \\
    \textbf{0.20} & \sub{0.31} & \sub{--}   & \sub{0.41} & \sub{0.48} & \sub{0.56} & \sub{0.62} \\
    \textbf{0.40} & \sub{0.34} & \sub{0.59} & \sub{--}   & \sub{0.57} & \sub{0.63} & \sub{0.68} \\
    \textbf{0.60} & \sub{0.36} & \sub{0.52} & \sub{0.43} & \sub{--}   & \sub{0.54} & \sub{0.62} \\
    \textbf{0.80} & \sub{0.33} & \sub{0.44} & \sub{0.37} & \sub{0.46} & \sub{--}   & \sub{0.60} \\
    \textbf{1.00} & \sub{0.28} & \sub{0.38} & \sub{0.32} & \sub{0.38} & \sub{0.40} & \sub{--}   \\
    \bottomrule
  \end{tabular}
\end{table}

\subsubsection{Numerical bound violations and irrigation-bound activity}

\begin{figure}[!htbp]
    \centering
    \includegraphics[width=0.9\linewidth]{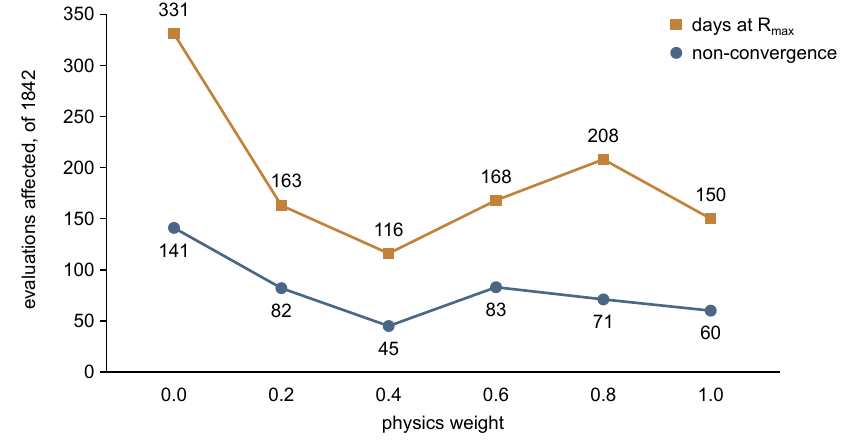}
    \caption{Number of affected runs out of 1842 (614 validation cases, three seeds). The `non-convergence' curve (bound violation) counts runs with at least one bound violation in the final iteration; the `days at $R_{\mathrm{max}}$' curve counts runs with at least one day at the upper bound. Both curves count runs with at least one event, not total hours or days, so they cannot be converted into the win fractions of \Cref{tab:lambda_nonconv,tab:lambda_rmax}.}
    \label{fig:lambda_rmax}
\end{figure}

Two further diagnostics describe how prediction and constrained control interact. For each run, the control code counts the hours with a bound violation in the final iteration and the days at the upper irrigation bound. \Cref{tab:lambda_nonconv,tab:lambda_rmax} compare these counts pair by pair. \Cref{fig:lambda_rmax} gives a different view: how many runs had at least one event. The two views are related, but one cannot be calculated from the other.

\begin{table}[H]
  \centering
  \caption{Physics-weight comparison on numerical bound violations, over 614 validation cases and three seeds. Each fraction rests on 54--118 non-tied pairs (107 for $\lambda=0.4$ against $\lambda=0$). Entries $\geq 0.75$ are in bold.}
  \label{tab:lambda_nonconv}
  \footnotesize
  \setlength{\tabcolsep}{7pt}
  \renewcommand{\arraystretch}{1.25}
  \begin{tabular}{@{} l c c c c c c @{}}
    & \multicolumn{6}{c}{$\lambda_B$} \\
    \cmidrule(l){2-7}
    $\lambda_A$ & \textbf{0.00} & \textbf{0.20} & \textbf{0.40} & \textbf{0.60} & \textbf{0.80} & \textbf{1.00} \\
    \midrule
    \textbf{0.00} & \sub{--}   & \sub{0.17} & \sub{0.06} & \sub{0.15} & \sub{0.19} & \sub{0.13} \\
    \textbf{0.20} & \textbf{0.83}       & \sub{--}   & \sub{0.21} & \sub{0.48} & \sub{0.44} & \sub{0.35} \\
    \textbf{0.40} & \textbf{0.94}       & \textbf{0.79}       & \sub{--}   & \textbf{0.77}       & \sub{0.66} & \sub{0.59} \\
    \textbf{0.60} & \textbf{0.85}       & \sub{0.52} & \sub{0.23} & \sub{--}   & \sub{0.46} & \sub{0.39} \\
    \textbf{0.80} & \textbf{0.81}       & \sub{0.56} & \sub{0.34} & \sub{0.54} & \sub{--}   & \sub{0.39} \\
    \textbf{1.00} & \textbf{0.87}       & \sub{0.65} & \sub{0.41} & \sub{0.61} & \sub{0.61} & \sub{--}   \\
    \bottomrule
  \end{tabular}
\end{table}

\begin{table}[H]
  \centering
  \caption{Physics-weight comparison on use of the upper irrigation bound $R_{\mathrm{max}}$, over 614 validation cases and three seeds. Ties are excluded as in \Cref{eq:psup}; each fraction rests on 129--374 non-tied pairs (274 for $\lambda=0.4$ against $\lambda=0$). Entries $\geq 0.75$ are in bold.}
  \label{tab:lambda_rmax}
  \footnotesize
  \setlength{\tabcolsep}{7pt}
  \renewcommand{\arraystretch}{1.25}
  \begin{tabular}{@{} l c c c c c c @{}}
    & \multicolumn{6}{c}{$\lambda_B$} \\
    \cmidrule(l){2-7}
    $\lambda_A$ & \textbf{0.00} & \textbf{0.20} & \textbf{0.40} & \textbf{0.60} & \textbf{0.80} & \textbf{1.00} \\
    \midrule
    \textbf{0.00} & \sub{--}   & \sub{0.18} & \sub{0.11} & \sub{0.23} & \sub{0.33} & \sub{0.23} \\
    \textbf{0.20} & \textbf{0.82}       & \sub{--}   & \sub{0.31} & \sub{0.54} & \sub{0.57} & \sub{0.46} \\
    \textbf{0.40} & \textbf{0.89}       & \sub{0.69} & \sub{--}   & \sub{0.70} & \sub{0.69} & \sub{0.59} \\
    \textbf{0.60} & \textbf{0.77}       & \sub{0.46} & \sub{0.30} & \sub{--}   & \sub{0.56} & \sub{0.43} \\
    \textbf{0.80} & \sub{0.67} & \sub{0.43} & \sub{0.31} & \sub{0.44} & \sub{--}   & \sub{0.38} \\
    \textbf{1.00} & \textbf{0.77}       & \sub{0.54} & \sub{0.41} & \sub{0.57} & \sub{0.62} & \sub{--}   \\
    \bottomrule
  \end{tabular}
\end{table}

The selected $\lambda=0.4$ exceeds the descriptive 0.75 threshold against $\lambda=0$, 0.2 and 0.6 on numerical bound violations, and against $\lambda=0$ on upper-bound use. Together with the lower open-loop drift, these comparisons support the chosen hybrid weight, but no single weight is best on every metric: data-only training tracks slightly better in closed loop. The weaker results of physics-only training are consistent with the mismatch between the Gardner and VGM models.

\section{Multi-property extrapolation}
\label{app:extrap}

\begin{figure}[!htbp]
  \centering
  \includegraphics[width=0.82\linewidth]{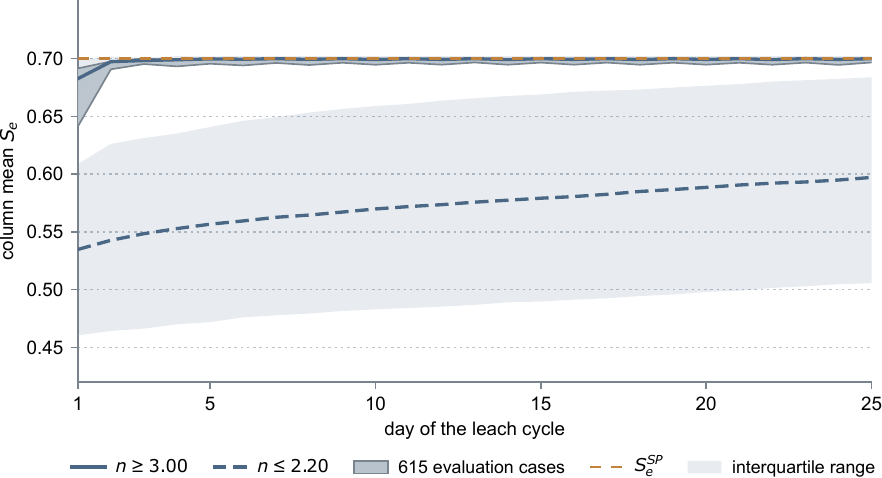}
  \caption{Spatial-mean effective saturation over the cycle when all four hydraulic parameters are moved 10\% of their range widths beyond their bounds, grouped by which bound $n$ crossed. Shaded bands show the middle half of the pooled runs, and the dashed line is the set-point. The extra curve shows the evaluation-set result for comparison.}
  \label{fig:extrap_multi}
\end{figure}

The joint test moves $K_s$, $\alpha$, $n$ and $\theta_s$ together beyond their upper or lower bounds by 10\% of each range width. Each of the $2^4=16$ combinations uses 64 initial saturations. Results are pooled by the direction of the $n$ excursion, which separates the two groups seen in \Cref{fig:extrap_multi}. Combinations with $n$ above its upper bound give a pooled final mean of 0.70, with an interquartile range of 0.01. Combinations with $n$ below its lower bound give 0.60 and 0.18 (\Cref{tab:extrap_multi}), a final deviation of 14.7\%, consistent with the single-parameter test.

\begin{table}[H]
  \centering
  \caption{Terminal spatial-mean effective saturation for cases in which all four hydraulic parameters were displaced beyond their bounds by 10\% of their respective training-range widths. Each group pools eight combinations, 512 cases and three seeds, separated by which bound $n$ crossed. The interquartile range is the width containing the middle half of the pooled runs.}
  \label{tab:extrap_multi}
  \footnotesize
  \setlength{\tabcolsep}{11pt}
  \renewcommand{\arraystretch}{1.25}
  \begin{tabular}{@{} l c c c @{}}
    \toprule
    \textbf{Group} & \textbf{Terminal mean $S_e$} & \textbf{Interquartile range} & \textbf{Deviation, \%} \\
    \midrule
    $n \leq 2.20$ & 0.60 & 0.18 & 14.7 \\
    $n \geq 3.00$ & 0.70 & 0.01 & \sub{0.0} \\
    \bottomrule
  \end{tabular}
\end{table}

\section{Robustness to measurement and parameter errors}
\label{app:stress}

\begin{figure}[!htbp]
  \centering
  \includegraphics[width=0.95\linewidth]{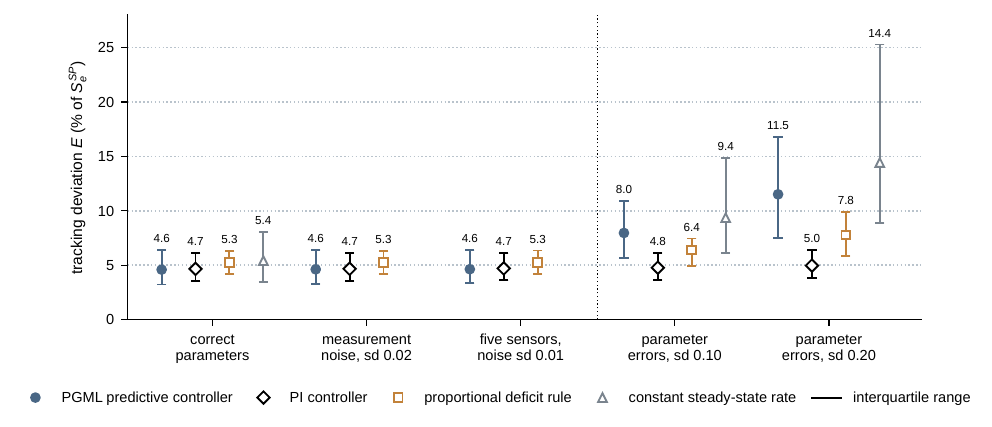}
  \caption{Robustness to measurement and parameter errors on 205 evaluation cases (seed-42 network). Markers give the median full-cycle RMS deviation $E$ and bars the interquartile range. The dotted vertical line separates the measurement tests from the parameter-error tests. The constant rule uses no measurements, so the noise and sensor tests do not apply to it.}
  \label{fig:robustness}
\end{figure}

\Cref{tab:robust,fig:robustness} compare the four controllers under the measurement and parameter-error conditions of \Cref{sec:protocol}. Measurement noise and sparse sensors changed the median deviations of all feedback controllers by less than 0.1 percentage points. Parameter errors had a much larger effect on the controllers that rely on the supplied parameters: the median PGML deviation rose from 4.6\% to 8.0\% and 11.5\% at levels 0.10 and 0.20, and the constant rule from 5.4\% to 9.4\% and 14.4\%, whereas PI changed only from 4.7\% to 4.8\% and 5.0\%, because its integral action removes persistent offsets. The final distance from the set-point shows the same pattern. These results should be read as a sensitivity test rather than a calibrated uncertainty analysis: the multiplicative factors can move parameters outside the training range and, for $n$, towards the admissible limit $n>2$, so the parameter-error rows combine parameter mismatch with extrapolation of the surrogate. An offset-free formulation or online parameter estimation would be needed for the predictive controller to match the robustness of PI.

\begin{table}[H]
  \centering
  \caption{Robustness to measurement and parameter errors on 205 evaluation cases (seed-42 network). Each entry gives the median full-cycle RMS deviation $E$ [\%] and, in brackets, the median distance of the final spatial-mean saturation from the set-point. Parameter errors multiply each of $K_s$, $\alpha$, $n$ and $\theta_s$ given to the controller by an independent log-normal factor at the reported level; the reference simulator keeps the true values. The constant rule uses no measurements, so noise and sensor tests do not apply to it.}
  \label{tab:robust}
  \footnotesize
  \setlength{\tabcolsep}{5pt}
  \renewcommand{\arraystretch}{1.2}
  \begin{tabular}{@{} l c c c c @{}}
    \toprule
    \textbf{Condition} & \textbf{PGML predictive} & \textbf{PI controller} & \textbf{prop.\ deficit rule} & \textbf{constant rate} \\
    \midrule
    correct parameters, full profile & 4.59 (0.002) & 4.66 (0.000) & 5.26 (0.000) & 5.42 (0.000) \\
    measurement noise, sd 0.02 & 4.62 (0.003) & 4.67 (0.001) & 5.26 (0.001) & -- \\
    five sensors, noise sd 0.01 & 4.63 (0.004) & 4.70 (0.003) & 5.26 (0.004) & -- \\
    parameter errors, level 0.10 & 7.96 (0.037) & 4.77 (0.000) & 6.40 (0.017) & 9.37 (0.042) \\
    parameter errors, level 0.20 & 11.52 (0.069) & 4.96 (0.000) & 7.77 (0.033) & 14.44 (0.088) \\
    \bottomrule
  \end{tabular}
\end{table}

\section{Nomenclature}

\noindent\begin{minipage}[t]{0.49\linewidth}
\small\textbf{Roman symbols}\\[2pt]
\begin{tabular}{@{}l@{\hspace{0.8em}}p{0.72\linewidth}@{}}
$D,\,D_G$      & moisture diffusivity, VGM and Gardner, cm$^2$/day \\
$e_d$          & saturation deficit at the start of day $d$, \Cref{eq:pdrule} \\
$E_{s,b}$      & full-cycle RMS tracking deviation of case $s$ at seed $b$, \% of set-point \\
$f$            & hourly surrogate map, \Cref{eq:onestep} \\
$H$            & column height, 550~cm \\
$h$            & pressure head, cm \\
$J$            & controller objective, \Cref{eq:objective} \\
$K,\,K_G$      & hydraulic conductivity, VGM and Gardner, cm/day \\
$K_s$          & saturated hydraulic conductivity, cm/day \\
$K_p,\,K_i$    & proportional and integral gains, \Cref{eq:pdrule,eq:pi} \\
$\mathcal L_d,\,\mathcal L_p,\,\mathcal L$ & data, physics and total training loss \\
$m$            & van Genuchten exponent, $m=1-2/n$ \\
$N_h$          & hourly steps per leach cycle, 600 \\
$N_R$          & candidate irrigation rates per update, 200 \\
$N_s$          & evaluation cases, 615 \\
$N_z$          & spatial intervals, 220 (nodes $i=0,\ldots,N_z$) \\
$n$            & van Genuchten shape parameter \\
$q,\,q_{i+1/2}$ & Darcy flux, cm/day \\
$q_f$          & target irrigation fraction of the operational cap \\
\end{tabular}
\end{minipage}\hfill
\begin{minipage}[t]{0.49\linewidth}
\small\textbf{Roman symbols (continued)}\\[2pt]
\begin{tabular}{@{}l@{\hspace{0.5em}}p{0.60\linewidth}@{}}
$R$            & irrigation rate, cm/day; $R^k$ held over step $k$, $R_d$ over day $d$ \\
$R^\star$      & selected irrigation rate, \Cref{eq:objective} \\
$\mathcal R$   & candidate grid of $N_R$ irrigation rates \\
$R_{\mathrm{min}},\,R_{\mathrm{max}}$ & bounds of $\mathcal R$, \Cref{eq:rate_bounds} \\
$R_{\mathrm{num}},\,R_{\mathrm{stable}}$ & numerical ceiling and last accepted rate \\
$R_{\mathrm{scale,min}},\,R_{\mathrm{scale,max}}$ & training-scaler limits on $R$ \\
$S_e$          & effective saturation, $\theta/\theta_s$ \\
$S_e^{SP}$     & saturation set-point, 0.70 \\
$\bar S_{e,d}$ & measured column-mean $S_e$ at the start of day $d$ \\
$S_{e,\mathrm{init}}$ & uniform initial column saturation \\
$s$            & case index \\
$t$            & continuous time, day \\
$z$            & depth below the surface, cm \\
\end{tabular}
\end{minipage}

\vspace{0.8em}
\noindent\begin{minipage}[t]{0.49\linewidth}
\small\textbf{Greek symbols}\\[2pt]
\begin{tabular}{@{}l@{\hspace{0.8em}}p{0.70\linewidth}@{}}
$\alpha$       & van Genuchten retention parameter, 1/cm \\
$\alpha_g,\,\beta$ & Gardner diffusivity and conductivity parameters \\
$\gamma$       & excursion fraction beyond a training bound \\
$\Delta t,\,\Delta z$ & time step, 1/24~day; node spacing, 2.5~cm \\
$\delta$       & VGM conductivity exponent, $3+2/(n-2)$ \\
$\theta,\,\theta_s$ & volumetric moisture content and its saturated value \\
$\lambda$      & physics weight in the training loss \\
$\sigma$       & log-amplitude of the irrigation excitation \\
\end{tabular}
\end{minipage}\hfill
\begin{minipage}[t]{0.49\linewidth}
\small\textbf{Indices and markers}\\[2pt]
\begin{tabular}{@{}l@{\hspace{0.8em}}p{0.70\linewidth}@{}}
$(\cdot)^k$    & hourly step, $k=1,\ldots,N_h$ \\
$(\cdot)^{\ell}$ & time level in the residual, $\ell\in\{k,k+1\}$ \\
$(\cdot)_i$    & spatial node, $i=0,\ldots,N_z$ \\
$(\cdot)_b$    & batch member (Appendix~\ref{app:residual}); training seed (on $E_{s,b}$) \\
$(\cdot)_d$    & day of the leach cycle \\
$(\cdot)_e$    & effective (on $S$) \\
$(\cdot)_G,\,(\cdot)_g$ & Gardner constitutive model \\
$(\cdot)_s$    & saturated (on $\theta$, $K$); case index when standalone \\
$\widehat{(\cdot)}$ & surrogate prediction; sample estimate on $p_{AB}$ \\
$\overline{(\cdot)}$ & mean \\
\end{tabular}
\end{minipage}

\vspace{0.5em}
\noindent\small\textbf{Abbreviations:} MPC, model predictive control; NMPC, nonlinear MPC; PGML, physics-guided machine learning; PI, proportional--integral; RMS(E), root-mean-square (error); SP, set-point; VGM, van Genuchten Modified.
\normalsize


\end{document}